\documentclass[a4paper,11pt]{article}
\usepackage{jheppub}
\usepackage{array}
\usepackage{amsmath,amssymb,latexsym}
\usepackage{bm}
\usepackage[svgnames]{xcolor}

\usepackage{graphicx}
\usepackage{float}
\usepackage{subcaption}

\usepackage{mathtools}
\usepackage{physics}
\usepackage{braket}

\usepackage{natbib}
\usepackage{graphicx}

\usepackage{shuffle} 
\usepackage{slashed} 

\usepackage{multirow}

\newcommand{\ccal}{\mathcal{C}}

\allowdisplaybreaks

\usepackage{amsthm}

\begin{document}

\title{A three-body form factor at sub-leading power in the high-energy limit: planar contributions}

\author{Yishuai Guo,}
\author{Zhi-Feng Liu,}
\author{Mingming Lu,}
\author{Tianya Xia,}
\author{Li Lin Yang}
\affiliation{Zhejiang Institute of Modern Physics, School of Physics, Zhejiang University, Hangzhou 310027, China}
\emailAdd{yishuaiguo@zju.edu.cn}
\emailAdd{xiangyuas@zju.edu.cn}
\emailAdd{mingming.lu@zju.edu.cn}
\emailAdd{xiatianya@zju.edu.cn}
\emailAdd{yanglilin@zju.edu.cn}

\abstract{
We consider two-loop planar contributions to a three-body form factor at the next-to-leading power in the high-energy limit, where the masses of external particles are much smaller than their energies. The calculation is performed by exploiting the differential equations of the expansion coefficients, both for facilitating the linear relations among them, and for deriving their analytic expressions. The result is written in terms of generalized polylogarithms involving a few simple symbol letters. Our method can be readily applied to the calculation of non-planar contributions as well. The result provides crucial information for establishing sub-leading factorization theorems for massive scattering amplitudes in the high-energy limit.
}

\maketitle
                     
\newpage

\section{Introduction}

The Standard Model (SM) of particle physics contain a couple of massive particles at the electroweak scale, including the Higgs boson, the top quark and the electroweak gauge bosons. An important quest of particle physics is to study the properties of these particles at the energy and luminosity frontiers with high precision, and probe new physics beyond the SM. This requires high precision theoretical predictions for the production processes involving these massive particles. However, the calculation of scattering amplitudes involving massive particles is generally much more difficult than the calculation of massless scattering amplitudes. On the other hand, due to the high energies of the Large Hadron Collider (LHC) and future colliders, the masses of the particles are often small compared to other kinematic invariants in the scattering processes. With such scale hierarchies, the perturbative scattering amplitudes and cross sections can develop large logarithms involving the ratios between the small masses and the large kinematic invariants in the high energy limit.

The standard way to deal with scale hierarchies is the method of factorization. Such factorization can be organized into different orders in an expansion parameter $\lambda \sim m^2/E^2$, where $m$ denotes the low scale of the small masses, while $E$ denotes the high scale of other kinematic invariants. The leading power (LP) corresponds to order $\lambda^0$. The factorization formula at the LP has been well-understood \cite{Penin:2005eh, Mitov:2006xs, Becher:2007cu, Gritschacher:2013pha, Hoang:2014tha, Liu:2017axv, Blumlein:2018tmz, Engel:2018fsb, Broggio:2022htr, Wang:2023qbf}. It has been shown in \cite{Wang:2023qbf} that a massive scattering amplitude in the high energy limit can be factorized into a massless amplitude, a soft function and several collinear functions (one for each external leg). Such a factorization formula can be used to predict the structure of the large logarithms of the form $\ln(m/E)$ at higher orders in perturbation theory, and also allows the resummation of these logarithms to all orders.

Given the importance of the factorization approach, it is essential to investigate the behavior of the massive amplitudes at sub-leading powers in $\lambda$. On one hand, this allows us to learn about the typical size of power corrections, and therefore provides an uncertainty estimate of calculations based on the LP factorization formula. On the other hand, once we understand the factorization structure at sub-leading powers, we can extrapolate the high-energy approximations to intermediate energy ranges, and eventually combine them with low-energy approximations based on threshold/soft factorization. This will lead to an adequate description of the scattering amplitudes in the whole phase space.

During the past years, studies of small-mass factorization at the next-to-leading power (NLP) in $\lambda$ have emerged in the literature~\cite{Penin:2014msa, Larkoski:2014bxa, Moult:2015aoa, Kolodrubetz:2016uim, Feige:2017zci, Liu:2017vkm, Beneke:2017ztn, Gervais:2017yxv, Beneke:2018rbh, Liu:2018czl, Liu:2019oav, Laenen:2020nrt, Liu:2020eqe, Liu:2020tzd, Liu:2020wbn, Liu:2022ajh, terHoeve:2023ehm, vanBijleveld:2025ekz}, either within the framework of soft-collinear effective theory (SCET) \cite{Bauer:2000ew,Bauer:2000yr,Bauer:2001ct,Bauer:2001yt,Beneke:2002ph,Beneke:2002ni,Hill:2002vw} or based on diagrammatic analysis. The analysis of sub-leading soft emissions has been carried out in \cite{Larkoski:2014bxa}. The construction of an easy-to-use helicity operator basis has been studied in \cite{Moult:2015aoa, Kolodrubetz:2016uim, Feige:2017zci}. Anomalous dimensions of sub-leading power $N$-jet operators have been calculated in \cite{Beneke:2017ztn,Beneke:2018rbh}. Based on power counting and region analysis, the NLP factorization formula in the small-mass limit has been proposed for the Yukawa theory \cite{Gervais:2017yxv} and for quantum electrodynamics (QED) \cite{Laenen:2020nrt, terHoeve:2023ehm}. An important ingredient in the QED factorization formula, the NLP jet function, has recently been computed in \cite{vanBijleveld:2025ekz}. However, the NLP factorization formula has not been generically proven.

An important validation of the factorization formula is to compare its fixed-order expansion with a direct computation of some massive scattering amplitudes or form factors in the small-mass limit. The $1 \to 2$ massive quark form factor has been computed at two loops \cite{Bernreuther:2004ih,Gluza:2009yy,Blumlein:2020jrf}, and three-loop calculations are in progress \cite{Henn:2016kjz,Grozin:2017aty,Ablinger:2017hst,Lee:2018nxa,Lee:2018rgs,Ablinger:2018yae,Blumlein:2018tmz,Blumlein:2019oas,Fael:2022rgm,Fael:2022miw,Fael:2023zqr,Blumlein:2023uuq}. However, $1 \to 2$ kinematics does not capture all possible structure at the two-loop order and beyond. In this paper, we initiate a study of a two-loop $1 \to 3$ massive form factor at sub-leading powers. At two loops, non-trivial correlations can only occur among at most 3 external particles. Therefore, the structure extracted from $1 \to 3$ form factors is generic enough to be applied to other form factors and scattering amplitudes. To this end, our ultimate goal is to study the $Q\bar{Q}g$ form factor in quantum chromodynamics (QCD), where $Q$ represents a massive quark, and $g$ is the gluon. Due to the complexity of this problem, we will begin with a simpler one in QED, i.e., the $e^+e^-\gamma$ form factor.

The scattering amplitudes involving massive electrons in QED are phenomenologically interesting on their own. In the future plans of high-energy physics experiments, various high-energy and high-luminosity $e^+e^-$ colliders have been proposed \cite{ILC:2013jhg, ILDConceptGroup:2020sfq, Bambade:2019fyw, CEPCStudyGroup:2018ghi, TLEPDesignStudyWorkingGroup:2013myl, FCC:2018byv, FCC:2018evy}. On these machines, it is important to understand the standard QED processes as precise as possible. These processes provide crucial inputs for the calibration of the detectors, and for a better understanding of beam parameters. The studies in this work are therefore useful for these applications.

The remainder of the paper is organized as follows: we introduce the notation in Section \ref{sec:setup}, and describe the method to determine independent coefficients in the small-mass expansion in Section \ref{sec:coefficients}. In Section \ref{sec:differential}, we demonstrate the solution of the differential equations satisfied by the independent coefficients, and discuss the final results. We briefly summarize in Section \ref{sec:summary} where an outlook for future works is also described.

\section{Setup of the calculation}
\label{sec:setup}

We consider the $1 \to 3$ QED process
\begin{equation}
        \label{eq:process}
	\gamma^{*}(p_4) \rightarrow e^+(p_1) + e^-(p_2) + \gamma(p_3).
\end{equation}
The independent kinematic variables are
\begin{equation}
        \label{eq:kinemactic_varibales}
        s_{12} = (p_1+p_2)^2 \,, \quad
	s_{23} = (p_2+p_3)^2 \,, \quad
	s_{123} = (p_1+p_2+p_3)^2 \,, \quad
	p_1^2 = p_2^2 = m^2 \,,
\end{equation}
where $m$ is the mass of the electron. For convenience, we define the following dimensionless variables
\begin{equation}
	\label{eq:dimensionless_kinematic_variables}
	x = \frac{m^2}{-s_{123}} \,, \quad
	y = \frac{s_{12}}{s_{123}} \,, \quad
	z = \frac{s_{23}}{s_{123}} \,.
\end{equation}

The results for the tree-level amplitude $\mathcal{M}^{(0)}$ and the one-loop amplitude $\mathcal{M}^{(1)}$ have been well-established \cite{Ellis:1980wv, Fabricius:1981sx}. In this work, we are concerned with the two-loop amplitude $ \mathcal{M}^{(2)}$. We generate the relevant two-loop diagrams and the corresponding amplitudes using \texttt{QGRAF} \cite{Nogueira:1991ex}. These diagrams can be categorized into eight integral families: five planar and three non-planar. We only consider the planar families in this work. The amplitudes containing Lorentz and Dirac indices can be decomposed into an appropriate basis with scalar coefficients (form factors). For simplicity, in this work we demonstrate our method using the the interference term in the squared-amplitude $\mathcal{F}^{(2)} = \langle \mathcal{M}^{(0)} | \mathcal{M}^{(2)} \rangle$ where the indices are summed over. We use \texttt{FeynCalc} \cite{Mertig:1990an, Shtabovenko:2016sxi, Shtabovenko:2020gxv, Shtabovenko:2023idz} and \texttt{FORM} \cite{Vermaseren:1992vn, Vermaseren:2000nd} to manipulate the expressions, and express the results in terms of two-loop scalar integrals. These scalar integrals can be expressed as
\begin{equation}
	\label{eq:family}=-
	I_{\{a_i\}}
    \equiv e^{2 \epsilon \gamma_{E}} \left(-s_{123}\right)^{a-d}  \int\frac{\dd^d k_1}{i \pi^{\frac{d}{2}}}\frac{\dd^d k_2}{i \pi^{\frac{d}{2}}} \prod_{i=1}^9 \frac{1}{D_i^{a_i}} \,,
\end{equation}
where $a = \sum_i a_i$, and we have multiplied an appropriate power of $(-s_{123})$ to make the scalar integrals dimensionless. 
Each set $\{D_i\}$ defines an integral family. 
There are 5 planar families involved in this work. The corresponding $\{D_i\}$ are given by
\begin{subequations}
\begin{align}
    \label{eq:topo_1}\{&k_1^2-m^2, (k_1-p_1)^2, (k_1-p_1-p_2)^2-m^2, (k_2-p_1-p_2)^2-m^2, (k_2-p_1-p_2-p_3)^2-m^2,\notag\\
	& k_2^2-m^2, (k_1-k_2)^2, (k_1-p_1-p_3)^2, (k_2-p_1)^2\} \,,
    \\
    \label{eq:topo_2}\{&k_1^2 - m^2, k_2^2, (k_1-k_2)^2-m^2, (k_1-p_1)^2,(k_1-p_1-p_2)^2-m^2,  (k_1-p_1-p_2-p_3)^2-m^2,\notag
    \\
	& (k_2-p_1)^2-m^2, (k_2-p_1-p_2)^2, (k_2-p_1-p_2-p_3)^2\} \,,
    \\
    \label{eq:topo_3}\{&k_1^2-m^2, k_2^2, (k_1-k_2)^2-m^2, (k_1-p_2)^2, (k_1-p_2-p_3)^2,(k_1-p_1-p_2-p_3)^2-m^2,\notag
    \\
	& (k_2-p_2)^2-m^2, (k_2-p_2-p_3)^2-m^2, (k_2-p_1-p_2-p_3)^2\} \,,
    \\
    \label{eq:topo_4}\{&k_1^2-m^2, k_2^2, (k_1-k_2)^2, (k_1-p_1)^2, (k_1-p_1-p_2)^2-m^2,(k_1-p_1-p_2-p_3)^2-m^2,\notag
    \\
	&(k_2-p_1)^2, (k_2-p_1-p_2)^2-m^2, (k_2-p_1-p_2-p_3)^2-m^2\} \,,
    \\
    \label{eq:topo_5}\{&k_1^2-m^2, k_2^2-m^2, (k_1-k_2)^2, (k_1-p_1)^2, (k_1-p_1-p_2)^2-m^2,(k_1-p_1-p_2-p_3)^2-m^2,\notag
    \\
	& (k_2-p_1)^2, (k_2-p_1-p_2)^2, (k_2-p_1-p_2-p_3)^2-m^2\} \,.
\end{align}
\end{subequations}
Note that an integral family may contain several topologies, where a topology is defined according to which $D_i$'s appear in the denominator of Eq.~\eqref{eq:family} (i.e., $a_i > 0$). The topologies belonging to each of the 5 families are illustrated in Fig.~\ref{fig:diagramg} of Appendix~\ref{appendix:topology definition}.

For each family, we use \texttt{Kira} \cite{Maierhofer:2017gsa, Klappert:2020nbg} to reduce the scalar integrals to a set of MIs $\mathcal{I}_i(x,y,z)$ by solving integration-by-parts (IBP) relations. The contribution from this family to the two-loop squared-amplitude can then be decomposed as
\begin{equation}
	\label{eq:amplitude}
    \mathcal{F}^{(2)} \ni \left( \frac{\mu^2}{-s_{123}} \right)^{2\epsilon} \sum_{i} \mathcal{A}_{i}(\epsilon,x,y,z) \, \mathcal{I}_{i}(x,y,z) \,,
\end{equation}
where $\mu$ is the renormalization scale, while the coefficients $\mathcal{A}_{i}$ are rational functions, and can be easily expanded in the limit of small $x$. The remaining task is then to compute the master integrals in the high-energy limit $x \to 0$, while keeping the exact dependence on the variables $y$ and $z$. To illustrate our method, we focus on the family \eqref{eq:topo_2}. There are 123 MIs in this family. In the high-energy limit, these MIs admit the asymptotic expansion
\begin{equation}
	\label{eq:series_expansion}
	\mathcal{I}_{i}(x,y,z) = \sum_{n_1=-2}^{\infty} \sum_{n_2=0}^{\infty} \sum_{n_3=0}^{n_{3,\max}} \mathcal{C}_{i,n_1,n_2,n_3}(y,z) \, \epsilon^{n_1} \, x^{n_2} \, \log^{n_3} (x) \,.
\end{equation}
The purpose of this work then reduces to the calculation of the coefficients as a function of $y$ and $z$.
Note that the two-loop integrals considered in this work have no soft-collinear overlapped divergences, and the minimal power of $\epsilon$ is $-2$. The integrals also have no power-like singularities in the limit $m \to 0$, and therefore the minimal power of $x$ is $0$. The maximal power of $\log(x)$, $n_{3,\max}$, depends on the value of $n_1$ and $n_2$. In practice, we will truncate the series in $\epsilon$ and $x$ for the purpose of calculating the squared-amplitude $\mathcal{F}^{(2)}$ up to $x^1$ and $\epsilon^0$. This corresponds to the next-to-leading power (NLP) in the high-energy expansion. Nevertheless, our method can be used to compute higher power terms as well.

\section{Differential equations and independent coefficients}
\label{sec:coefficients}

To compute the coefficients, we adopt the method of differential equations \cite{Kotikov:1990kg,Remiddi:1997ny,Gehrmann:1999as}. From the IBP reduction, we can construct the differential equations of the MIs $\mathcal{I}_i$ with respect to the variable $t \in \{x,y,z\}$:
\begin{equation}\label{eq:differential_equation_matrix}
	\frac{\partial}{\partial t}
	\begin{pmatrix}
		\mathcal{I}_1 \\
		\vdots \\
		\mathcal{I}_{123}
	\end{pmatrix} =
	\begin{pmatrix}
		\mathcal{P}^{(t)}_{1,1} & \cdots & \mathcal{P}^{(t)}_{1,123} \\
		\vdots            & \ddots & \vdots \\
		\mathcal{P}^{(t)}_{123,1}& \cdots & \mathcal{P}^{(t)}_{123,123}
	\end{pmatrix}
	\begin{pmatrix}
		\mathcal{I}_1 \\
		\vdots \\
		\mathcal{I}_{123}
	\end{pmatrix}
    \,,
\end{equation}
where $\mathcal{P}^{(t)}_{i,j}$ are rational functions of $\epsilon$, $x$, $y$ and $z$. The differential equations with respect to $x$ lead to relations among the coefficients. Therefore, we only need to compute a set of independent coefficients, which is conceptually similar to the master integrals. After determining the set of independent coefficients, we can employ their differential equations with respect to $y$ and $z$ to obtain their analytic expressions.

\subsection{Determination of expansion orders}
\label{sec:exp_ord}

We apply the expansion \eqref{eq:series_expansion} into the differential equations \eqref{eq:differential_equation_matrix} with $t = x$. We note that
\begin{equation}
    \frac{\partial}{\partial x} x^{n_2} \log^{n_3}(x) = x^{n_2-1} \left[ n_2 \log^{n_3}(x) + n_3 \log^{n_3-1}(x) \right] .
\end{equation}
That is, taking derivative with respect to $x$ always decreases the power of $x$, but may or may not decrease the power of $\log(x)$. On the other hand, the matrix elements $\mathcal{P}^{(x)}_{i,j}$ may also contain poles at $x = 0$. Therefore, the differential equations with respect to $x$ lead to linear relations among the coefficients $\mathcal{C}_{i,n_1,n_2,n_3}(y,z)$. From these linear relations, we can determine a set of independent coefficients. We will refer to these independent coefficients as ``master coefficients'' (MCs), and express the remaining coefficients as linear combinations of the MCs. 

Before solving the linear relations (there are an infinite number of them), we need to constrain the highest values of the integers $n_1$ and $n_2$ for each $i$, i.e, the maximal powers of $\epsilon$ and $x$ in the expansion of the master integral $\mathcal{I}_i$ (note that the maximal power of $\log(x)$ is naturally determined given the values of $n_1$ and $n_2$). The first constraint comes from the required orders in the expansion of the squared-amplitude $\mathcal{F}^{(2)}$. As we have mentioned, we truncate the expansion in $x$ up to the NLP, i.e., order $x^1$. In addition, we truncate the expansion in $\epsilon$ up to order $\epsilon^0$. These requirements impose constraints on $n_1$ and $n_2$ for each master integral $\mathcal{I}_i$. However, we find that these constraints are too tight for our purpose: the MCs determined from these constrained linear relations are not closed under differentiation. The reason is that there are cancellations among different topologies in a family when their contributions go into Eq.~\eqref{eq:amplitude}.

In order to obtain a closed system of differential equations, we need to slightly loosen the constraints. For that purpose, we expand the coefficients in the differential equations as well:
\begin{equation}
	\label{eq:series_expansion_Pij}
	\mathcal{P}^{(t)}_{i,j} = \sum_{k,l} \mathcal{B}^{(t)}_{i,j,k,l}(y,z) \epsilon^{k}x^{l} \,.
\end{equation}
The lowest values of $k$ and $l$ are then crucial for the determination of the highest expansion orders.
As an example, we consider the differential equations with respect to $x$. The constraints from $y$ and $z$ can be similarly studied. Plugging Eqs.~\eqref{eq:series_expansion} and \eqref{eq:series_expansion_Pij} into \eqref{eq:differential_equation_matrix}, we find that
\begin{equation}
    \label{eq:MI_m2_relation_simplify}
    n_2 \, \mathcal{C}_{i,n_1,n_2,n_3}(y,z) + (n_3+1) \, \mathcal{C}_{i,n_1,n_2,n_3+1}(y,z)=\sum_{j,k,l} \mathcal{B}^{(x)}_{i,j,k,l}(y,z) \, \mathcal{C}_{j,n_1-k,n_2-1-l,n_3}(y,z) \,.
\end{equation}

This is the master equation for relations among expansion coefficients. For this system to be closed, it is necessary that all coefficients contributing to a given $(n_1,n_2,n_3)$ are incorporated in the above equation.

Let's focus on the equation with $i=11$, which arises from the differential equation of $\mathcal{I}_{11}$ with respect to $x$. In this case, one has $j \in \{1,8,11\}$ on the right-hand side of Eq.~\eqref{eq:MI_m2_relation_simplify}. For the NLP accuracy of the squared amplitude, one requires both $\mathcal{I}_{11}$ and $\mathcal{I}_{1}$ to be expanded to order $x^3$, since the minimal power of $x$ in $\mathcal{A}_1$ and $\mathcal{A}_{11}$ is $-2$, which means that we need to consider the equation with $n_2 = 3$. However, the minimal value of $l$ for $i=11$, $j=1$ is $l=-2$. Therefore, the coefficient with $n_2-1-l=4$ will appear on the right-hand side of the equation. For this reason, we need to expand $\mathcal{I}_{1}$ up to order $x^4$.

Considerations similar to the above also applied to the constrains on $n_1$, i.e., the expansion orders in $\epsilon$. We need to perform such analysis for each sector, and update the constraints until all equations are closed. As a final outcome of such analyses, we provide the number of MCs for each integral family in Table~\ref{tab:numMCs}.
\begin{table}[!htbp]
    \label{tab:MCEnumration}
    \centering
    \begin{tabular}{|l|c|c|c|c|c|}
    \hline
    Integral families & $F_1$ & $F_2$ & $F_3$ & $F_4$ & $F_5$ \\ \hline
    Number of MCs     & 247     & 359     & 383     & 158     & 199     \\ \hline
    \end{tabular}
    \caption{The number of MCs in each integral family.}
    \label{tab:numMCs}
\end{table}

\subsection{Solving the linear relations among coefficients}

After obtaining a system of linear equations of the coefficients, we can solve the equations to express all coefficients in terms of the MCs. Before doing that, we need to decide a set of rules for choosing the MCs. This is similar to the rules for choosing the MIs in Feynman integral reduction. In practice, this requires to determine between two coefficients which one is preferred over the other.

The choice of MCs will greatly influence the process of solving their differential equations with respect to $y$ and $z$. We would like them to have as simple analytic expressions as possible. From experience, this then corresponds to selecting coefficients with smaller $n_1$, smaller $n_2$ and larger $n_3$. Coefficients in lower sectors are also preferred over those in higher sectors.

After deciding the preference, we can solve the linear system using the ``user-defined system'' functionality of \texttt{Kira}. We can then construct the system of differential equations of the MCs. For that we need to re-express the derivatives of the MCs in terms of the MCs again. This step can be accomplished by \texttt{Kira} as well. For example, we write the differential equation of a MC with respect to $y$ as
\begin{equation}
	\label{eq:diff_equ_y}
	\frac{\partial}{\partial y} \mathcal{C}_{i,n_1,n_2,n_3}(y,z) =
	\sum_{j,k,l} \mathcal{B}^{(y)}_{i,j,k,l}(y,z) \, \mathcal{C}_{j,n_1-k,n_2-l,n_3}(y,z) \,.
\end{equation}
The above equation can be provided to $\texttt{Kira}$ along with other linear relations. One then obtains the left-hand side expressed in terms of MCs. Finally, we arrive at the differential equations
\begin{align}
	\label{eq:diff_equ_indep_coeff}
	\frac{\partial}{\partial y} \mathcal{C}_{I}(y,z) = \sum_{J} \mathcal{A}^{(y)}_{I,J}(y,z) \, \mathcal{C}_{J}(y,z) \,, \nonumber
    \\
	\frac{\partial}{\partial z} \mathcal{C}_{I}(y,z) = \sum_{J} \mathcal{A}^{(z)}_{I,J}(y,z) \, \mathcal{C}_{J}(y,z) \,,
\end{align}
where $I=(i,n_1,n_2,n_3)$ and $J=(j,m_1,m_2,m_3)$ denote two subscript sequences that label the MCs.

\section{Solving the differential equations}
\label{sec:differential}

After constructing the differential equations, we now want to find the analytic solutions for the MCs. Ideally, the solution procedure can be greatly simplified if the connection matrices $\mathcal{A}^{(y)}_{I,J}(y,z)$ and $\mathcal{A}^{(z)}_{I,J}(y,z)$ take a strictly triangular form. In that case the solutions can be easily written in terms of iterated integrals. This is similar in spirit to the idea of canonical differential equations \cite{Henn:2013pwa}. However, the existing tools \cite{Gituliar:2017vzm,Prausa:2017ltv,Meyer:2017joq,Lee:2020zfb,Dlapa:2020cwj} for finding canonical differential equations do not easily adapt to the current case. Fortunately, the connection matrices are already in a block-triangular form with at most $3 \times 3$ blocks. These can be iteratively converted to a strictly triangular form, as will be demonstrated in the following.

\subsection{Simple cases}
\label{section:self-coupled_coefficient}
We first consider the simple cases where the derivatives of a MC only depend on itself and the already-solved MCs:
\begin{equation}
	\label{eq:self-coupled}
	\begin{aligned}
		\frac{\partial}{\partial y} \mathcal{C}(y,z) &=\mathcal{A}_{y}(y,z) \, \mathcal{C}(y,z) + \mathcal{G}_{y}(y,z) \,,
        \\
		\frac{\partial}{\partial z} \mathcal{C}(y,z) &=\mathcal{A}_{z}(y,z) \, \mathcal{C}(y,z) + \mathcal{G}_{z}(y,z) \,,
	\end{aligned}
\end{equation}
where $\mathcal{G}_{y}$ and $\mathcal{G}_{z}$ come from the already-solved MCs. If the coefficients $\mathcal{A}_{y}$ and $\mathcal{A}_{z}$ are both zero, the above equations can be easily solved by direct integration. This corresponds to a strictly triangular system discussed earlier. We now consider the case where $\mathcal{A}_{y}$ and $\mathcal{A}_{z}$ are nonzero.

For convenience, we introduce the following functions:
\begin{equation}
    \mathcal{M}_{y}(y,z) = \exp \left(-\int \mathcal{A}_{y}(y,z) \, \mathrm{d}y \right) , \quad \mathcal{M}_{z}(y,z) = \exp \left(-\int \mathcal{A}_{z}(y,z) \, \mathrm{d}z  \right) .
\end{equation}
Multiplying the MC by the two functions, the differential equations Eq.~\eqref{eq:self-coupled} are transformed into:
\begin{equation}
	\label{eq:self-coupled_trans}
	\frac{\partial}{\partial y} \left( \mathcal{M}_y\mathcal{C}(y,z)\right)  =\mathcal{M}_y \mathcal{G}_{y}(y,z) \,,\quad
	\frac{\partial}{\partial z} \left( \mathcal{M}_z\mathcal{C}(y,z)\right)  =\mathcal{M}_z \mathcal{G}_{z}(y,z) \,.
\end{equation}
One can see that only already-solved MCs appear on the right-hand side of the above two equations. We would like to find a transformation function $\mathcal{T}(y,z)$ such that 
\begin{equation}
	\label{eq:self-coupled-zero}
	\frac{\partial}{\partial y} \left( \mathcal{T}\mathcal{C}(y,z)\right)  =\mathcal{T} \mathcal{G}_{y}(y,z) \, ,\quad
	\frac{\partial}{\partial z} \left( \mathcal{T}\mathcal{C}(y,z)\right)  =\mathcal{T} \mathcal{G}_{z}(y,z) \, .
\end{equation}
This is an example of a strictly triangular system of differential equations, where the derivatives of the transformed MCs only depend on already-solved ones. Comparing Eq.~\eqref{eq:self-coupled-zero} with Eq.~\eqref{eq:self-coupled_trans}, one can see that
\begin{equation}
    \mathcal{T}(y,z) = \mathcal{M}_{y}(y,z) \, f_z(z) = \mathcal{M}_{z}(y,z) \, f_y(y) \,,
\end{equation}
where the function $f_z(z)$ satisfies the differential equation
\begin{equation}
    \frac{\dd}{\dd z} \ln f_z(z) = \frac{\partial}{\partial z} \ln \frac{\mathcal{M}_{z}(y,z)}{\mathcal{M}_{y}(y,z)} \,,
\end{equation}
and similar for $f_y(y)$. The fact that the right-hand side of the above equation is independent of $y$ follows from the compatibility condition of Eq.~\eqref{eq:self-coupled}:
\begin{equation}
    \frac{\partial}{\partial y} \frac{\partial}{\partial z} \ln \frac{\mathcal{M}_z}{\mathcal{M}_y} = \partial_z \mathcal{A}_{y} - \partial_y \mathcal{A}_{z} = 0 \,,
\end{equation}
which further follows from the linear-independence of $\mathcal{C}(y,z)$ with respect to $\mathcal{G}_y$, $\mathcal{G}_z$ and their derivatives. From the above analysis, we can write
\begin{equation}
    \label{eq:transform_T}
    \mathcal{T}(y,z) = \mathcal{M}_y(y,z) \, \exp\left( \int \dd z  \, \frac{\partial}{\partial z} \ln \frac{\mathcal{M}_{z}(y,z)}{\mathcal{M}_{y}(y,z)} \right) .
\end{equation} 

Building on this result, we get general solution to the differential equaiton \eqref{eq:self-coupled-zero}:
\begin{equation}
	\mathcal{C}(y,z) = \frac{1}{\mathcal{T}} \left[ \int \mathcal{T} \mathcal{G}_{y} \dd y + \mathcal{C}_{z}(z) \right]
	= \frac{1}{\mathcal{T}} \left[ \int \mathcal{T} \mathcal{G}_{z} \dd z + \mathcal{C}_{y}(y) \right] ,
\end{equation}
where $\mathcal{C}_{y}(y)$ is a function that only depends on $y$, and similarly for $\mathcal{C}_{z}(z)$. These two functions are related by the second equal sign in the above formula, and are only fixed once the arbitrary constant terms of the indefinite integrals are chosen. Using the fact that $\mathcal{C}_{y}(y)$ is independent of $z$, we can derive the differential equation of $\mathcal{C}_{z}(z)$ with respect to $z$:
\begin{equation}
	\frac{\partial \mathcal{C}_{z}}{\partial z} = \mathcal{T} \mathcal{G}_{z} - \frac{\partial}{\partial z} \int \mathcal{T} \mathcal{G}_{y} \dd y  \, .
\end{equation}
which allows us to determine the analytic expression of $\mathcal{C}_{z}$ (as well as $\mathcal{C}_{y}$) up to a constant of integration. This constant should be fixed by a boundary condition, to be discussed later.

The above procedure of transforming the differential equations to a strictly triangular form can be naturally extended to $2 \times 2$ blocks and $3 \times 3$ blocks, as will be discussed in the following subsections.

\subsection{$2\times2$ blocks}
\label{section:2x2_coupled_coefficient}

We now consider the $2 \times 2$ blocks, where the differential equations take the form
\begin{equation}
	\partial_t
	\begin{bmatrix}
		\ccal_1 \\
		\ccal_2
	\end{bmatrix} =
	\begin{bmatrix}
		\mathcal{A}_{11}^{(t)} & \mathcal{A}_{12}^{(t)} \\
		\mathcal{A}_{21}^{(t)} & \mathcal{A}_{22}^{(t)}
	\end{bmatrix}
	\begin{bmatrix}
		\ccal_1 \\
		\ccal_2
	\end{bmatrix} + 
	\begin{bmatrix}
		\mathcal{G}_1^{(t)} \\
		\mathcal{G}_2^{(t)}
	\end{bmatrix}.
\end{equation}
where $t \in \{y, z\}$, $\mathcal{A}^{(t)}_{ij}$'s are rational functions of $y$ and $z$, and $\mathcal{G}^{(t)}_i$'s come from already-solved MCs. To solve the equations, we would like to find a linear combination $\ccal' = b_1 \ccal_1 + b_2 \ccal_2$ with $b_1$ and $b_2$ being rational functions of $y$ and $z$, such that the derivatives of $\ccal'$ only depend on itself. Taking the derivatives, one has
\begin{equation}
	\partial_t {\ccal'} = \frac{\partial_t b_{1}+b_{1} \mathcal{A}_{11}^{(t)}+ b_{2} \mathcal{A}_{21}^{(t)}}{b_{1}}b_{1}\ccal_1+ \frac{\partial_t b_{2}+b_{1} \mathcal{A}_{12}^{(t)}+ b_{2} \mathcal{A}_{22}^{(t)}}{b_{2}}b_{2}\ccal_2 + b_{1} \mathcal{G}_1^{(t)}+ b_{2} \mathcal{G}_2^{(t)} \,.
\end{equation}
We therefore require
\begin{equation}
	\label{eq:b1_and_b2_relation}
	\mathcal{A}'_t \equiv \frac{\partial_t b_{1}+b_{1} \mathcal{A}_{11}^{(t)}+ b_{2} \mathcal{A}_{21}^{(t)}}{b_{1}} = \frac{\partial_t b_{2}+b_{1} \mathcal{A}_{12}^{(t)}+ b_{2} \mathcal{A}_{22}^{(t)}}{b_{2}} \,.
\end{equation}

Once we find a suitable pair of $b_1$ and $b_2$, the differential equation of $\ccal'$ becomes
\begin{equation}
	\partial_t \ccal' = \mathcal{A}'_t \ccal' + \mathcal{G}'_t \,, \quad \mathcal{G}'_t = b_1 \mathcal{G}_1^{(t)} + b_2 \mathcal{G}_2^{(t)} \,.
\end{equation}
The above equation resembles the one studied in Section~\ref{section:self-coupled_coefficient}. We can employ the method there to get rid of the term $\mathcal{A}'_t \ccal'$, and effectively transform the differential equation to a strictly triangular form. After obtaining the solution of $\ccal'$, we can use it to derive analytic expressions of $\ccal_1$ and $\ccal_2$. Taking $\ccal_1$ as an example, we have
\begin{equation}
	\partial_t \ccal_1=\mathcal{A}_{11}^{(t)}\ccal_1+\mathcal{A}_{12}^{(t)}\ccal_2+\mathcal{G}_1^{(t)}
	=\left(\mathcal{A}_{11}^{(t)}-\frac{\mathcal{A}_{12}^{(t)}b_1}{b_2}\right)\ccal_1+\frac{\mathcal{A}_{12}^{(t)}}{b_{2}} \ccal' +\mathcal{G}_1^{(t)} \,.
\end{equation}
Since $\ccal'$ is known, the above equation can again be solved using the method of Section~\ref{section:self-coupled_coefficient}. The expression of $\ccal_2$ can then be obtained from those of $\ccal'$ and $\ccal_1$.

From the above discussions, it is clear that the task boils down to finding a particular solution of $b_1$ and $b_2$ satisfying Eq.~\eqref{eq:b1_and_b2_relation}. Note that Eq.~\eqref{eq:b1_and_b2_relation} is actually a differential equation for the ratio $b_2/b_1$, which can be rewritten as
\begin{equation}
    \partial_t \left( \frac{b_2}{b_1} \right) - \mathcal{A}_{21}^{(t)} \left( \frac{b_2}{b_1} \right)^2 + \left( \mathcal{A}_{22}^{(t)} - \mathcal{A}_{11}^{(t)} \right) \frac{b_2}{b_1} + \mathcal{A}_{12}^{(t)} = 0 \,.
\end{equation}
This is the so-called Riccati equation. There are two ways to find a solution.

\paragraph{Method one: brute-force solution.}
For the cases we encounter in this work, the Riccati equation can be solved with the help of \texttt{Mathematica}. As an example, we may need to solve the system with
\begin{equation}
    \mathcal{A}^{(y)} =
	\begin{pmatrix}
		-\dfrac{y+z-1}{y (y+2 z-1)} & \dfrac{y-1}{y^2 (y+2 z-1)}\\
		\dfrac{y (z+1)-(z-1)^2}{(y-1) (y+2 z-1)} & \dfrac{-y (3z+2)-3z+2}{(y-1) y (y+2 z-1)}
	\end{pmatrix} .
\end{equation}
For convenience, we can set $b_1 = 1$, and the Riccati equation with respect to $y$ reads
\begin{equation}
    \frac{\partial}{\partial y} b_2 - \frac{y(z+1)-(z-1)^2}{(y-1)(y+2 z-1)} \, b_2^2+\frac{3+y^2-4 z-2 y(2+z)}{(y-1)y(y+2 z-1)} \, b_2 + \frac{y-1}{y^2(y+2 z-1)} =0 \,.
\end{equation}
Using \texttt{Mathematica}, we can obtain a general solution of the above equation, which involves arbitrary functions of $z$. We require that the solution is rational, and also satisfies the corresponding equation with respect to $z$. This give rises to a particular solution
\begin{equation}
    b_2=\frac{(y-1)^2}{y(y z+y+z-1)} \,.
\end{equation}
In practice, this method is sometimes slow, and we may resort to the second method.

\paragraph{Method two: solution by ansatz.} Since the equation only constrains $b_2/b_1$, we can assume that both $b_1$ and $b_2$ are polynomials of $y$ and $z$ with integer coefficients. We can then make the ansatz
\begin{equation}
		b_{1}=\sum_{i=0}^{n}\sum_{j=0}^{n-i} c_{ij}^{(1)} y^i z^j \,, \quad
		b_{2}=\sum_{i=0}^{n}\sum_{j=0}^{n-i} c_{ij}^{(2)} y^i z^j \,,
\end{equation}
where the coefficients $c_{ij}^{(1)}$ and $c_{ij}^{(2)}$, as well as the degree $n$, are yet to be determined.

We start with a small value of $n$, and substitute the ansatz into Eq.~\eqref{eq:b1_and_b2_relation}. This then leads to a system of quadratic equations of the coefficients. If the system of equations only allows a trivial solution (i.e., all coefficients being zero), it means that the value of $n$ is too small. We then increase the value of $n$ until find a non-trivial solution.

For the example considered above, we find that a non-trivial solution can be obtained with $n=3$:
\begin{equation}
	b_1=-y (y z+y+z-1) \,, \quad
	b_2=-(y-1)^2 \,,
\end{equation}
which is equivalent to the result of the first method.

\subsection{$3\times3$ blocks}

Similar to the $2\times2$ cases, the differential equations for $3\times3$ blocks have the form
\begin{equation}
	\label{eq:3x3_coupled}
	\partial_t
	\begin{pmatrix}
		\ccal_1 \\
		\ccal_2 \\
		\ccal_3
	\end{pmatrix} =
	\begin{pmatrix}
		\mathcal{A}^{(t)}_{11} & \mathcal{A}^{(t)}_{12} & \mathcal{A}^{(t)}_{13} \\
		\mathcal{A}^{(t)}_{21} & \mathcal{A}^{(t)}_{22} & \mathcal{A}^{(t)}_{23} \\
		\mathcal{A}^{(t)}_{31} & \mathcal{A}^{(t)}_{32} & \mathcal{A}^{(t)}_{33} \\
	\end{pmatrix}
	\begin{pmatrix}
		\ccal_1 \\
		\ccal_2 \\
		\ccal_3
	\end{pmatrix} + 
	\begin{pmatrix}
		\mathcal{G}^{(t)}_1 \\
		\mathcal{G}^{(t)}_2 \\
		\mathcal{G}^{(t)}_3
	\end{pmatrix}
    .
\end{equation}
The method for dealing with this system is analogous. First, we need to construct a linear combination $\ccal' = b_1 \ccal_1 + b_2 \ccal_2 + b_3 \ccal_3$, whose differential equations only involve itself. This boils down to the following equation
\begin{align}
	\label{eq:b1_b2_and_b3_relation}
	\mathcal{A}' &\equiv \frac{\partial_t b_1+b_1 \mathcal{A}^{(t)}_{11} + b_2 \mathcal{A}^{(t)}_{21}+b_3 \mathcal{A}^{(t)}_{31}}{b_1} \nonumber
    \\
    &= \frac{\partial_t b_2+ b_1 \mathcal{A}^{(t)}_{12}+ b_2 \mathcal{A}^{(t)}_{22}+b_3 \mathcal{A}^{(t)}_{32}}{b_2} \nonumber
    \\
    &=\frac{\partial_t b_3+ b_1 \mathcal{A}^{(t)}_{13}+ b_2 \mathcal{A}^{(t)}_{23}+b_3 \mathcal{A}^{(t)}_{33}}{b_3} \,.
\end{align}
This equation can be rewritten as
\begin{align}
    \partial_{t} \left(\frac{b_2}{b_1}\right) - \mathcal{A}^{(t)}_{21}\left(\frac{b_2}{b_1}\right)^2+\left(\mathcal{A}^{(t)}_{22}-\mathcal{A}^{(t)}_{11}-\mathcal{A}^{(t)}_{31} \frac{b_3}{b_1} \right)  \frac{b_2}{b_1}+\mathcal{A}^{(t)}_{32} \frac{b_3}{b_1} +\mathcal{A}^{(t)}_{12} &= 0 \,, \nonumber
    \\
    \partial_{t} \left(\frac{b_3}{b_1}\right) - \mathcal{A}^{(t)}_{31}\left(\frac{b_3}{b_1}\right)^2+\left(\mathcal{A}^{(t)}_{33}-\mathcal{A}^{(t)}_{11}-\mathcal{A}^{(t)}_{21}\frac{b_2}{b_1}\right)\frac{b_3}{b_1}+\mathcal{A}^{(t)}_{23}\frac{b_2}{b_1}+\mathcal{A}^{(t)}_{13} &= 0 \,.
\end{align}
One can observe that the two unknowns $b_2/b_1$ and $b_3/b_1$ are still coupled by the two equations, and this system is challenging to solve by brute force. Fortunately, the second method introduced in the previous subsection still works, and offers an algorithmic way to tackle this kind of problems. We make polynomial ansatz for $b_1$, $b_2$ and $b_3$, and solve for the coefficients from Eq.~\eqref{eq:b1_b2_and_b3_relation}. After that, we can solve for $\ccal'$ using the method of Section~\ref{section:self-coupled_coefficient}. We are then left with a $2\times2$ system of $\ccal_1$ and $\ccal_2$, which can be solved following the strategy of Section~\ref{section:2x2_coupled_coefficient}.

Let's again demonstrate our approach using an example, where
\begin{equation}
    \mathcal{A}^{(y)} =
	\begin{pmatrix}
		\frac{1}{1-y} & -\frac{y+2 z-1}{(y-1) y (y+z-1)} &
		\frac{z-1}{(y-1) y (y+z-1)} \\
		\frac{y (z+1)}{(y-1) (y+z)} & \frac{y \left(4 y z+y+3
			z^2-3 z-1\right)+(1-z) z}{(y-1) y \left(y^2+y (2
			z-1)+(z-1) z\right)} & -\frac{(z-1) (2 y+z-1)}{(y-1)
			\left(y^2+y (2 z-1)+(z-1) z\right)} \\
		\frac{z}{y+z} & \frac{z (3 y+2 z-2)}{y \left(y^2+y (2
			z-1)+(z-1) z\right)} & \frac{1-z}{y^2+y (2 z-1)+(z-1)
			z} \\
	\end{pmatrix}
    .
\end{equation}
A particular solution can be found by using the ansatz with $n=2$, and is given by
\begin{equation}
		b_1=y z \,, \quad
		b_2=2 z \,, \quad
		b_3=y-z \,.
\end{equation}
This give rise to a decoupled differential equation
\begin{equation}
	\frac{\partial \mathcal{C}'}{\partial y}=\frac{1}{y-1}\mathcal{C}'+y z \, \mathcal{G}^{(y)}_1+2z \, \mathcal{G}^{(y)}_2+(y-1) \, \mathcal{G}^{(y)}_3 \,.
\end{equation}
We don't go into the details of the remaining steps here.

In this work, we do not encounter blocks more than $3 \times 3$. But as one can see, the method introduced above can be applied to more complicated cases as well. This proceeds in an iterative way: one first solves for a linear combination whose differential equation is decoupled, which then leads to a simpler system.

\subsection{Boundary conditions and final results}

To obtain the solution to the system of differential equations, we still need to determine the boundary conditions for each coefficient. For that we use \texttt{AMFlow} \cite{Liu:2022chg} to numerically compute the small-mass expansion of all MIs at a specific kinematic point. During this process, all variables (except $m^2$ but including $\epsilon$), are set to exact values (rational numbers). For each value of $\epsilon$, the output has the following structure
\begin{equation}
        \label{eq:amflow_output}
        \sum_{i} \sum_{j=0}^{j_{\max}} w_{ij}(\epsilon) \left( m^2 \right)^{j + \alpha_i(\epsilon)} \,,
\end{equation}
where $\alpha_i(\epsilon)$ is a linear function of $\epsilon$, $w_{ij}(\epsilon)$ depends on $\epsilon$ and the kinematic variables (which we have suppressed), and $j_{\max}$ is specified according the required expansion order (in $m^2$) of the MIs. By varying $\epsilon$ with fixed kinematic variables, we can reconstruct the coefficients in $\alpha_i(\epsilon)$, as well as the function $w_{ij}(\epsilon)$ as a power series in $\epsilon$:
\begin{equation}
        \label{eq:amflow_ansatz}
        w_{ij}(\epsilon)= \sum_{k=k_{\min}}^{k_{\max}} a_{ijk} \, \epsilon^{k} \,.
\end{equation}
To determine the coefficients $a_{ijk}$, we need to sample at least $(k_{\max}-k_{\min}+1)$ different values of $\epsilon$. Note that the precision of the reconstructed $a_{ijk}$ also depends on the number of samples. Therefore, we need to choose an appropriate number according to the required expansion order (in $\epsilon$) of the MIs and the required precision for the coefficients. In practice, this number is about $50$.

With the reconstructed coefficients, we can reexpand Eq.~\eqref{eq:amflow_output} in $\epsilon$, and obtain the small-mass expansion of the MIs at the chosen kinematic point in the form
\begin{equation}
    \label{eq:amflow_final}
	\mathcal{I}_i(\epsilon,x) = \sum_{n_1,n_2,n_3} f_{i,n_1,n_2,n_3} \, \epsilon^{n_1} x^{n_2} \log^{n_3}(x) \,.
\end{equation}
The minimal powers of $\epsilon$ and $x$, and the maximal powers of $\log(x)$ for a given pair $(n_1,n_2)$ can be read off from the above expansion. Note that $f_{i,n_1,n_2,n_3}$ is just the coefficient $\ccal_{i,n_1,n_2,n_3}(y,z)$ at a particular kinematic point, and the general solution of $\ccal_{i,n_1,n_2,n_3}(y,z)$ has already been obtained from the differential equations. Such a solution can be written in terms of generalized polylogarithms (GPLs) with the help of \texttt{PolyLogTools} \cite{Duhr:2019tlz}, up to some unknown constants. These constants can then be reconstructed as transcendental numbers by the PSLQ algorithm, using the high-precision numbers (about 100 decimal digits) of $f_{i,n_1,n_2,n_3}$. The transcendental basis is generated from the elements $\{\pi,\ln(2),\zeta_2,\zeta_3,\zeta_4,\mathrm{Li}_4(1/2),\zeta_5\}$ up to weight 5. It is worth mentioning that the DEs may demand higher order coefficients in $\epsilon$, as discussed in Sec.~\ref{sec:exp_ord}. That's the reason why weight-5 functions and constants appear, despite the fact that two-loop amplitudes up to order $\epsilon^0$ only require weight-4 at most. Note that the form of the analytic expressions can be different in different kinematic regions. We concretely work in the Euclidean region where all planar MIs have no imaginary part. The expressions in other kinematic regions can be obtained via analytic continuation. Since our results are written in terms of GPLs, their analytic structure is very well understood. The branch cuts of GPLs are uniquely fixed by their symbols, and whenever one analytic continues across a branch cut, an extra term proportional to $\pm i\pi$ is generated, where the sign of $i\pi$ depends on the direction of going across the cut. See, e.g., \cite{Goncharov:2001iea, Vollinga:2004sn, zbMATH05494656} for more detailed discussions.

With the analytic expressions for the MIs, we can combine them to obtain the planar contributions to the squared-amplitude $\mathcal{F}^{(2)}$ up to order $x^1$ and $\epsilon^0$. It can be written in terms of GPLs up to weight $5$, with the following symbol letters:
\begin{equation}
    1-y,\, y,\, 1-2 z,\, 1-2 y-2 z,\, 1-z,\, 2-z,\, 1-y-z,\, z,\, 1+z,\, y+z,\, 1+y+z .
\end{equation}
There are two additional letters $\{1-y,\,2-y-z\}$ appearing in the MIs. But they cancel out in the final expression for the squared-amplitude. Note that the original massive amplitude involves elliptic integrals, but the expanded amplitude can be expressed by GPLs up to the NLP. The result for the squared-amplitude is very compact, with a size of about 400~KB. We attach the expression as an electronic file to this paper.

We have performed several sanity checks on our calculation. We have applied our method to the one-loop amplitude where the complete result can be easily obtained. Upon expansion in the high-energy limit, the complete result agrees exactly with our calculation. We have also numerically computed the MIs at several different kinematic points other than the chosen boundary point, and find that the results agree with the outcomes from our analytic expressions. In particular, we have verified kinematic points beyond the Euclidean region, and find that the imaginary parts agree as well. These checks demonstrate the reliability of our method.

\section{Summary and outlook}
\label{sec:summary}

In this work, we initiate a study of the sub-leading power contributions to the multi-parton massive form factors in the high-energy limit, where the parton masses are much smaller than their energies. These form factors provide crucial information to formulate and validate sub-leading power factorization theorems for generic scattering amplitudes. Such factorization theorems are important to resum the mass logarithms beyond the leading power, and to generate approximate results for scattering amplitudes at higher loops.

While the two-loop massive quark form factors are available, the $1 \to 2$ kinematics is not generic enough to be extended to multi-parton scattering amplitudes. Therefore, we focus on $1 \to 3$ form factors where two-loop multi-parton correlations can be studied. To establish the calculational techniques, we start with the two-loop planar contributions to a QED form factor $\gamma^* \to e^+e^-\gamma$ in this work. Although only two-particle correlations are present in this case, it is enough to demonstrate that our method is capable to obtain analytic expressions of the relevant loop integrals as a small-mass expansion.

Our calculations are based on the differential equations satisfied by the expansion coefficients of the master integrals. They can be derived up to an arbitrary order in the power expansion using the differential equations satisfied by the master integrals. The differential equations with respect to the mass $m$ are then utilized to derive linear relations among the expansion coefficients. Similar to the IBP reduction, these linear relations can be solve to express all coefficients in terms of a finite set of master coefficients. The differential equations of the master coefficients with respect to momenta invariants are then employed to solve for their analytic expressions. The solutions are expressed by GPLs, which are then combined and lead to a compact analytic result for the planar contributions to the form factor.

There are several obvious follow-ups to this work in sight. A rigorous verification of factorization at NLP necessitates the inclusion of non-planar contributions to the amplitudes. Our method can be readily applied to non-planar contributions to this form factor. The main obstacle in such a calculation lies in the IBP reduction. We have attempted with the state-of-art reduction tools such as \texttt{Kira}, \texttt{NeatIBP} \cite{Wu:2023upw, Wu:2025aeg} and \texttt{Blade} \cite{Guan:2024byi}, but have not succeeded within a reasonable amount of time. It will be necessary to develop more efficient reduction methods for such cutting-edge calculations. Another obvious generalization is to consider the $Q\bar{Q}g$ form factor in QCD, where genuine three-parton correlations can occur. This involves more integral families, but the method established in this work can still be straightforwardly applied. Finally, the differential equations for the master coefficients are solved by brute-force in this work. It will be interesting to investigate whether these differential equations can be casted into a canonical form which can be solved directly as iterated integrals. We leave these investigations to future works.

\section*{Acknowledgments}
This work was supported in part by the National Natural Science Foundation of China under Grant No. 12375097, 12347103, and the Fundamental Research Funds for the Central Universities.
Zhi-Feng Liu was supported by the China Postdoctoral Science Foundation
(2023M733123, 2023TQ0282) and the Postdoctoral Fellowship Program of China Postdoctoral
Science Foundation. 

\appendix
\section{Details of the integral families}
\label{appendix:topology definition}

\begin{figure}[t!]
    \centering
    \begin{subfigure}[t]{0.3\textwidth}
        \centering
        \includegraphics[width=\linewidth]{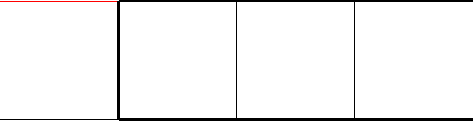}
        \caption{$F_1$ topology 1.}
    \end{subfigure}
    \begin{subfigure}[t]{0.3\textwidth}
        \centering
        \includegraphics[width=\linewidth]{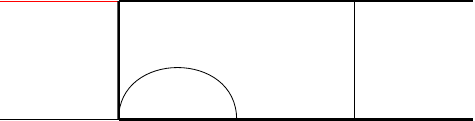}
        \caption{$F_1$ topology 2.}
    \end{subfigure}
    \begin{subfigure}[t]{0.3\textwidth}
        \centering
        \includegraphics[width=\linewidth]{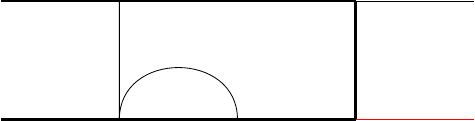}
        \caption{$F_1$ topology 3.}
    \end{subfigure}
    \begin{subfigure}[t]{0.3\textwidth}
        \centering
        \includegraphics[width=\linewidth]{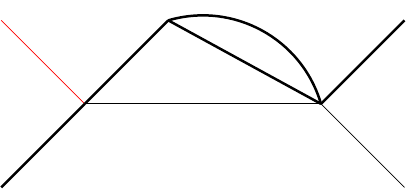}
        \caption{$F_1$ topology 4.}
    \end{subfigure}
    \begin{subfigure}[t]{0.3\textwidth}
        \centering
        \includegraphics[width=\linewidth]{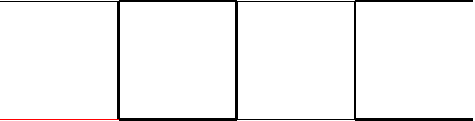}
        \caption{$F_2$ topology 1.}
    \end{subfigure}
    \begin{subfigure}[t]{0.3\textwidth}
        \centering
        \includegraphics[width=\linewidth]{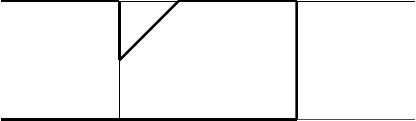}
        \caption{$F_2$ topology 2.}
    \end{subfigure}
    \begin{subfigure}[t]{0.3\textwidth}
        \centering
        \includegraphics[width=\linewidth]{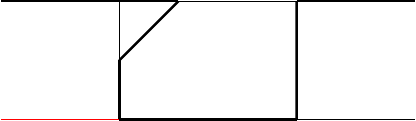}
        \caption{$F_2$ topology 3.}
    \end{subfigure}
    \begin{subfigure}[t]{0.3\textwidth}
        \centering
        \includegraphics[width=\linewidth]{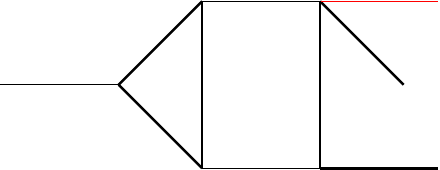}
        \caption{$F_2$ topology 4.}
    \end{subfigure}
    \begin{subfigure}[t]{0.3\textwidth}
        \centering
        \includegraphics[width=\linewidth]{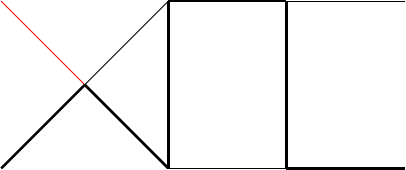}
        \caption{$F_2$ topology 5.}
    \end{subfigure}
    \begin{subfigure}[t]{0.3\textwidth}
        \centering
        \includegraphics[width=\linewidth]{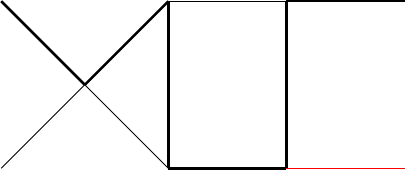}
        \caption{$F_2$ topology 6.}
    \end{subfigure}
    \begin{subfigure}[t]{0.3\textwidth}
        \centering
        \includegraphics[width=\linewidth]{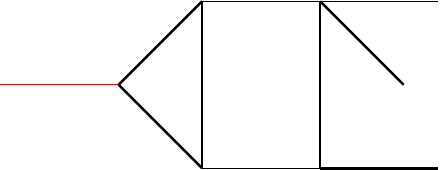}
        \caption{$F_2$ topology 7.}
    \end{subfigure}
    \begin{subfigure}[t]{0.3\textwidth}
        \centering
        \includegraphics[width=\linewidth]{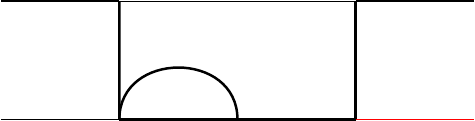}
        \caption{$F_2$ topology 8.}
    \end{subfigure}
    \begin{subfigure}[t]{0.3\textwidth}
        \centering
        \includegraphics[width=\linewidth]{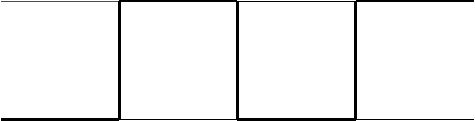}
        \caption{$F_3$ topology 1.}
    \end{subfigure}
    \begin{subfigure}[t]{0.3\textwidth}
        \centering
        \includegraphics[width=\linewidth]{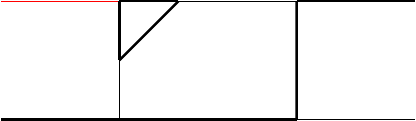}
        \caption{$F_3$ topology 2.}
    \end{subfigure}
    \begin{subfigure}[t]{0.3\textwidth}
        \centering
        \includegraphics[width=\linewidth]{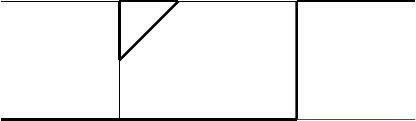}
        \caption{$F_3$ topology 3.}
    \end{subfigure}
    \begin{subfigure}[t]{0.3\textwidth}
        \centering
        \includegraphics[width=\linewidth]{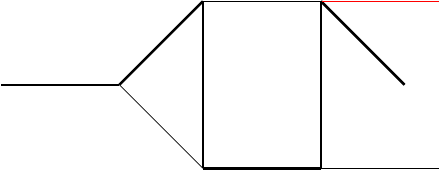}
        \caption{$F_3$ topology 4.}
    \end{subfigure}
    \begin{subfigure}[t]{0.3\textwidth}
        \centering
        \includegraphics[width=\linewidth]{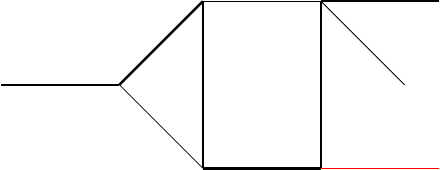}
        \caption{$F_3$ topology 5.}
    \end{subfigure}
    \begin{subfigure}[t]{0.3\textwidth}
        \centering
        \includegraphics[width=\linewidth]{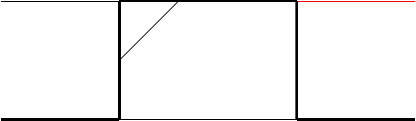}
        \caption{$F_4$ topology 1.}
    \end{subfigure}
    \begin{subfigure}[t]{0.3\textwidth}
        \centering
        \includegraphics[width=\linewidth]{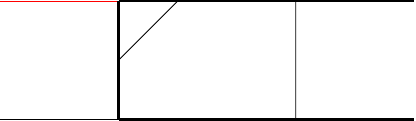}
        \caption{$F_5$ topology 1.}
    \end{subfigure}
    \caption{Two-loop planar topologies. Thick lines represent propagators with mass $m$ or external legs with $p^2 = m^2$. Thin lines represent massless propagators or external legs. Red lines represent external legs with $p^2 = s_{123}$.} 
    \label{fig:diagramg}
\end{figure}

As mentioned in the main text, we have 5 integral families, which contains 19 planar topologies in total, as shown in Fig.~\ref{fig:diagramg}. The MIs in each integral family are listed as follows:
\begin{align}
    &F_1\left(1,0,0,1,0,0,0,0,0\right), &&F_1\left(1,0,1,1,0,0,0,0,0\right), &&F_1\left(1,0,1,1,0,1,0,0,0\right),\notag\\
    &F_1\left(1,0,0,0,1,1,0,0,0\right), &&F_1\left(1,0,1,0,1,1,0,0,0\right), &&F_1\left(1,0,0,1,1,1,0,0,0\right),\notag\\
    &F_1\left(1,0,1,1,1,1,0,0,0\right), &&F_1\left(1,0,0,1,0,0,1,0,0\right), &&F_1\left(1,-1,0,1,0,0,1,0,0\right),\notag\\
    &F_1\left(0,1,0,1,0,0,1,0,0\right), &&F_1\left(1,1,0,1,0,0,1,0,0\right), &&F_1\left(1,0,0,0,1,0,1,0,0\right),\notag\\
    &F_1\left(1,-1,0,0,1,0,1,0,0\right), &&F_1\left(0,1,0,0,1,0,1,0,0\right), &&F_1\left(-1,1,0,0,1,0,1,0,0\right),\notag\\
    &F_1\left(1,1,0,0,1,0,1,0,0\right), &&F_1\left(1,1,-1,0,1,0,1,0,0\right), &&F_1\left(1,0,1,0,1,0,1,0,0\right),\notag\\
    &F_1\left(1,-1,1,0,1,0,1,0,0\right), &&F_1\left(1,0,1,-1,1,0,1,0,0\right), &&F_1\left(0,1,1,0,1,0,1,0,0\right),\notag\\
    &F_1\left(-1,1,1,0,1,0,1,0,0\right), &&F_1\left(1,1,1,0,1,0,1,0,0\right), &&F_1\left(1,1,1,-1,1,0,1,0,0\right),\notag\\
    &F_1\left(1,0,0,1,1,0,1,0,0\right), &&F_1\left(1,1,0,1,1,0,1,0,0\right), &&F_1\left(1,1,-1,1,1,0,1,0,0\right),\notag\\
    &F_1\left(1,1,0,1,1,-1,1,0,0\right), &&F_1\left(1,1,0,1,1,0,1,0,-1\right), &&F_1\left(1,0,1,1,1,0,1,0,0\right),\notag\\
    &F_1\left(1,-1,1,1,1,0,1,0,0\right), &&F_1\left(0,1,1,1,1,0,1,0,0\right), &&F_1\left(1,1,1,1,1,0,1,0,0\right),\notag\\
    &F_1\left(0,1,0,1,0,1,1,0,0\right), &&F_1\left(1,1,0,1,0,1,1,0,0\right), &&F_1\left(1,1,-1,1,0,1,1,0,0\right),\notag\\
    &F_1\left(0,1,0,0,1,1,1,0,0\right), &&F_1\left(-1,1,0,0,1,1,1,0,0\right), &&F_1\left(1,1,0,0,1,1,1,0,0\right),\notag\\
    &F_1\left(1,1,-1,0,1,1,1,0,0\right), &&F_1\left(1,1,0,-1,1,1,1,0,0\right), &&F_1\left(0,0,1,0,1,1,1,0,0\right),\notag\\
    &F_1\left(-1,0,1,0,1,1,1,0,0\right), &&F_1\left(0,-1,1,0,1,1,1,0,0\right), &&F_1\left(1,0,1,0,1,1,1,0,0\right),\notag\\
    &F_1\left(1,-1,1,0,1,1,1,0,0\right), &&F_1\left(1,0,1,-1,1,1,1,0,0\right), &&F_1\left(1,-2,1,0,1,1,1,0,0\right),\notag\\
    &F_1\left(0,1,1,0,1,1,1,0,0\right), &&F_1\left(-1,1,1,0,1,1,1,0,0\right), &&F_1\left(0,1,1,-1,1,1,1,0,0\right),\notag\\
    &F_1\left(0,1,1,0,1,1,1,-1,0\right), &&F_1\left(0,1,1,0,1,1,1,0,-1\right), &&F_1\left(-2,1,1,0,1,1,1,0,0\right),\notag\\
    &F_1\left(1,1,1,0,1,1,1,0,0\right), &&F_1\left(1,1,1,-1,1,1,1,0,0\right), &&F_1\left(1,1,1,0,1,1,1,-1,0\right),\notag\\
    &F_1\left(0,1,0,1,1,1,1,0,0\right), &&F_1\left(-1,1,0,1,1,1,1,0,0\right), &&F_1\left(1,1,0,1,1,1,1,0,0\right),\notag\\
    &F_1\left(1,1,-1,1,1,1,1,0,0\right), &&F_1\left(1,1,0,1,1,1,1,0,-1\right), &&F_1\left(0,1,1,1,1,1,1,0,0\right),\notag\\
    &F_1\left(-1,1,1,1,1,1,1,0,0\right), &&F_1\left(0,1,1,1,1,1,1,0,-1\right), &&F_1\left(1,1,1,1,1,1,1,0,0\right),\notag\\
    &F_1\left(1,1,1,1,1,1,1,-1,0\right), &&F_1\left(1,1,1,1,1,1,1,0,-1\right), &&F_1\left(1,0,0,1,0,0,0,1,0\right),\notag\\
    &F_1\left(0,0,0,0,0,1,1,1,0\right), &&F_1\left(-1,0,0,0,0,1,1,1,0\right), &&F_1\left(1,0,0,0,1,0,0,0,1\right),\notag\\
    &F_1\left(1,0,0,1,1,0,0,0,1\right), &&F_1\left(1,0,0,0,1,1,0,0,1\right), &&F_1\left(1,0,0,1,1,1,0,0,1\right),\notag\\
    &F_1\left(1,0,0,0,1,0,1,0,1\right), &&F_1\left(1,-1,0,0,1,0,1,0,1\right), &&F_1\left(1,0,-1,0,1,0,1,0,1\right),\notag\\
    &F_1\left(0,0,1,0,1,0,1,0,1\right), &&F_1\left(-1,0,1,0,1,0,1,0,1\right), &&F_1\left(0,-1,1,0,1,0,1,0,1\right),\notag\\
    &F_1\left(1,0,0,1,1,0,1,0,1\right), &&F_1\left(1,-1,0,1,1,0,1,0,1\right), &&F_1\left(1,0,-1,1,1,0,1,0,1\right),\notag\\
    &F_1\left(0,0,1,0,1,1,1,0,1\right), &&F_1\left(-1,0,1,0,1,1,1,0,1\right), &&F_1\left(0,-1,1,0,1,1,1,0,1\right)
\end{align}

\begin{align}
    &F_2\left(1,0,1,0,0,0,0,0,0\right), &&F_2\left(0,1,1,1,0,0,0,0,0\right), &&F_2\left(1,0,1,0,1,0,0,0,0\right),\notag\\
    &F_2\left(0,1,1,0,1,0,0,0,0\right), &&F_2\left(-1,1,1,0,1,0,0,0,0\right), &&F_2\left(0,1,1,1,1,0,0,0,0\right),\notag\\
    &F_2\left(1,0,1,0,0,1,0,0,0\right), &&F_2\left(0,1,1,0,0,1,0,0,0\right), &&F_2\left(-1,1,1,0,0,1,0,0,0\right),\notag\\
    &F_2\left(0,0,1,1,0,1,0,0,0\right), &&F_2\left(1,0,1,1,0,1,0,0,0\right), &&F_2\left(0,1,1,1,0,1,0,0,0\right),\notag\\
    &F_2\left(-1,1,1,1,0,1,0,0,0\right), &&F_2\left(0,1,1,1,0,1,-1,0,0\right), &&F_2\left(1,0,1,0,1,1,0,0,0\right),\notag\\
    &F_2\left(0,1,1,0,1,1,0,0,0\right), &&F_2\left(0,0,1,1,1,1,0,0,0\right), &&F_2\left(1,0,1,1,1,1,0,0,0\right),\notag\\
    &F_2\left(0,1,1,1,1,1,0,0,0\right), &&F_2\left(-1,1,1,1,1,1,0,0,0\right), &&F_2\left(0,1,1,1,1,1,-1,0,0\right),\notag\\
    &F_2\left(1,0,1,0,0,0,1,0,0\right), &&F_2\left(1,0,1,0,1,0,1,0,0\right), &&F_2\left(1,-1,1,0,1,0,1,0,0\right),\notag\\
    &F_2\left(0,1,1,0,1,0,1,0,0\right), &&F_2\left(-1,1,1,0,1,0,1,0,0\right), &&F_2\left(0,0,1,0,0,1,1,0,0\right),\notag\\
    &F_2\left(-2,0,1,0,0,1,1,0,0\right), &&F_2\left(1,0,1,0,0,1,1,0,0\right), &&F_2\left(1,-1,1,0,0,1,1,0,0\right),\notag\\
    &F_2\left(1,-2,1,0,0,1,1,0,0\right), &&F_2\left(0,1,1,0,0,1,1,0,0\right), &&F_2\left(-1,1,1,0,0,1,1,0,0\right),\notag\\
    &F_2\left(0,1,1,1,0,1,1,0,0\right), &&F_2\left(-1,1,1,1,0,1,1,0,0\right), &&F_2\left(0,0,1,0,1,1,1,0,0\right),\notag\\
    &F_2\left(1,0,1,0,1,1,1,0,0\right), &&F_2\left(1,-1,1,0,1,1,1,0,0\right), &&F_2\left(1,-2,1,0,1,1,1,0,0\right),\notag\\
    &F_2\left(0,1,1,0,1,1,1,0,0\right), &&F_2\left(-1,1,1,0,1,1,1,0,0\right), &&F_2\left(0,1,1,-1,1,1,1,0,0\right),\notag\\
    &F_2\left(0,1,1,0,1,1,1,-1,0\right), &&F_2\left(0,1,1,1,1,1,1,0,0\right), &&F_2\left(-1,1,1,1,1,1,1,0,0\right),\notag\\
    &F_2\left(1,1,0,0,0,0,0,1,0\right), &&F_2\left(1,1,0,0,1,0,0,1,0\right), &&F_2\left(1,1,1,0,1,0,0,1,0\right),\notag\\
    &F_2\left(1,1,0,0,0,1,0,1,0\right), &&F_2\left(1,0,1,0,0,1,0,1,0\right), &&F_2\left(1,-1,1,0,0,1,0,1,0\right),\notag\\
    &F_2\left(1,0,1,-1,0,1,0,1,0\right), &&F_2\left(0,1,1,0,0,1,0,1,0\right), &&F_2\left(-1,1,1,0,0,1,0,1,0\right),\notag\\
    &F_2\left(1,1,1,0,0,1,0,1,0\right), &&F_2\left(1,1,1,-1,0,1,0,1,0\right), &&F_2\left(1,1,1,0,0,1,-1,1,0\right),\notag\\
    &F_2\left(0,0,1,1,0,1,0,1,0\right), &&F_2\left(-1,0,1,1,0,1,0,1,0\right), &&F_2\left(0,-1,1,1,0,1,0,1,0\right),\notag\\
    &F_2\left(1,0,1,1,0,1,0,1,0\right), &&F_2\left(1,-1,1,1,0,1,0,1,0\right), &&F_2\left(1,0,1,1,-1,1,0,1,0\right),\notag\\
    &F_2\left(1,1,0,0,1,1,0,1,0\right), &&F_2\left(0,1,1,0,1,1,0,1,0\right), &&F_2\left(1,1,1,0,1,1,0,1,0\right),\notag\\
    &F_2\left(1,1,0,0,0,0,1,1,0\right), &&F_2\left(1,1,0,0,1,0,1,1,0\right), &&F_2\left(1,1,1,0,1,0,1,1,0\right),\notag\\
    &F_2\left(1,1,0,0,0,1,1,1,0\right), &&F_2\left(0,0,1,0,0,1,1,1,0\right), &&F_2\left(1,0,1,0,0,1,1,1,0\right),\notag\\
    &F_2\left(1,-1,1,0,0,1,1,1,0\right), &&F_2\left(1,0,1,-1,0,1,1,1,0\right), &&F_2\left(1,0,1,0,-1,1,1,1,0\right),\notag\\
    &F_2\left(1,0,1,0,0,1,1,1,-1\right), &&F_2\left(1,-2,1,0,0,1,1,1,0\right), &&F_2\left(1,-1,1,-1,0,1,1,1,0\right),\notag\\
    &F_2\left(0,1,1,0,0,1,1,1,0\right), &&F_2\left(-1,1,1,0,0,1,1,1,0\right), &&F_2\left(-2,1,1,0,0,1,1,1,0\right),\notag\\
    &F_2\left(1,1,1,0,0,1,1,1,0\right), &&F_2\left(1,1,1,-1,0,1,1,1,0\right), &&F_2\left(1,1,1,0,-1,1,1,1,0\right),\notag\\
    &F_2\left(0,0,1,1,0,1,1,1,0\right), &&F_2\left(-1,0,1,1,0,1,1,1,0\right), &&F_2\left(1,0,1,1,0,1,1,1,0\right),\notag\\
    &F_2\left(1,0,1,1,-1,1,1,1,0\right), &&F_2\left(1,1,0,0,1,1,1,1,0\right), &&F_2\left(0,1,1,0,1,1,1,1,0\right),\notag\\
    &F_2\left(0,1,1,0,1,1,1,1,-1\right), &&F_2\left(1,1,1,0,1,1,1,1,0\right), &&F_2\left(1,1,1,-1,1,1,1,1,0\right),\notag\\
    &F_2\left(1,1,1,0,1,1,1,1,-1\right), &&F_2\left(1,1,1,-2,1,1,1,1,0\right), &&F_2\left(1,1,1,-1,1,1,1,1,-1\right),\notag\\
    &F_2\left(0,0,1,1,0,0,0,0,1\right), &&F_2\left(-1,0,1,1,0,0,0,0,1\right), &&F_2\left(1,0,1,1,0,0,0,0,1\right),\notag\\
    &F_2\left(1,-1,1,1,0,0,0,0,1\right), &&F_2\left(1,0,1,0,1,0,0,0,1\right), &&F_2\left(1,-1,1,0,1,0,0,0,1\right),\notag\\
    &F_2\left(1,0,1,-1,1,0,0,0,1\right), &&F_2\left(0,0,1,1,1,0,0,0,1\right), &&F_2\left(-1,0,1,1,1,0,0,0,1\right),\notag\\
    &F_2\left(1,0,1,1,1,0,0,0,1\right), &&F_2\left(1,-1,1,1,1,0,0,0,1\right), &&F_2\left(1,0,1,0,0,0,1,0,1\right),\notag\\
    &F_2\left(1,-1,1,0,0,0,1,0,1\right), &&F_2\left(1,-2,1,0,0,0,1,0,1\right), &&F_2\left(0,0,1,0,1,0,1,0,1\right),\notag\\
    &F_2\left(-1,0,1,0,1,0,1,0,1\right), &&F_2\left(1,0,0,0,0,1,1,0,1\right), &&F_2\left(1,0,1,0,0,1,1,0,1\right),\notag\\
    &F_2\left(1,-1,1,0,0,1,1,0,1\right), &&F_2\left(1,0,1,-1,0,1,1,0,1\right), &&F_2\left(0,0,0,1,0,1,1,0,1\right),\notag\\
    &F_2\left(1,0,0,1,0,1,1,0,1\right), &&F_2\left(0,0,1,1,0,1,1,0,1\right), &&F_2\left(1,0,1,1,0,1,1,0,1\right),\notag\\
    &F_2\left(0,0,1,0,1,1,1,0,1\right), &&F_2\left(0,0,0,1,1,1,1,0,1\right), &&F_2\left(0,0,1,1,1,1,1,0,1\right)
\end{align}

\begin{align}
    &F_3\left(1,0,1,0,0,0,0,0,0\right), &&F_3\left(0,1,1,1,0,0,0,0,0\right), &&F_3\left(1,0,1,0,1,0,0,0,0\right),\notag\\
    &F_3\left(0,1,1,0,1,0,0,0,0\right), &&F_3\left(-1,1,1,0,1,0,0,0,0\right), &&F_3\left(1,0,1,0,0,1,0,0,0\right),\notag\\
    &F_3\left(0,1,1,0,0,1,0,0,0\right), &&F_3\left(-1,1,1,0,0,1,0,0,0\right), &&F_3\left(0,0,1,1,0,1,0,0,0\right),\notag\\
    &F_3\left(1,0,1,1,0,1,0,0,0\right), &&F_3\left(0,1,1,1,0,1,0,0,0\right), &&F_3\left(-1,1,1,1,0,1,0,0,0\right),\notag\\
    &F_3\left(0,1,1,1,0,1,-1,0,0\right), &&F_3\left(1,0,1,0,1,1,0,0,0\right), &&F_3\left(0,1,1,0,1,1,0,0,0\right),\notag\\
    &F_3\left(-1,1,1,0,1,1,0,0,0\right), &&F_3\left(1,0,1,0,0,0,1,0,0\right), &&F_3\left(1,0,1,0,1,0,1,0,0\right),\notag\\
    &F_3\left(1,-1,1,0,1,0,1,0,0\right), &&F_3\left(0,1,1,0,1,0,1,0,0\right), &&F_3\left(-1,1,1,0,1,0,1,0,0\right),\notag\\
    &F_3\left(0,0,1,0,0,1,1,0,0\right), &&F_3\left(-2,0,1,0,0,1,1,0,0\right), &&F_3\left(1,0,1,0,0,1,1,0,0\right),\notag\\
    &F_3\left(1,-1,1,0,0,1,1,0,0\right), &&F_3\left(1,-2,1,0,0,1,1,0,0\right), &&F_3\left(0,1,1,0,0,1,1,0,0\right),\notag\\
    &F_3\left(-1,1,1,0,0,1,1,0,0\right), &&F_3\left(0,1,1,1,0,1,1,0,0\right), &&F_3\left(-1,1,1,1,0,1,1,0,0\right),\notag\\
    &F_3\left(0,0,1,0,1,1,1,0,0\right), &&F_3\left(1,0,1,0,1,1,1,0,0\right), &&F_3\left(1,-1,1,0,1,1,1,0,0\right),\notag\\
    &F_3\left(1,-2,1,0,1,1,1,0,0\right), &&F_3\left(0,1,1,0,1,1,1,0,0\right), &&F_3\left(-1,1,1,0,1,1,1,0,0\right),\notag\\
    &F_3\left(0,1,1,-1,1,1,1,0,0\right), &&F_3\left(0,1,1,0,1,1,1,0,-1\right), &&F_3\left(-2,1,1,0,1,1,1,0,0\right),\notag\\
    &F_3\left(-1,1,1,-1,1,1,1,0,0\right), &&F_3\left(1,0,1,0,0,0,0,1,0\right), &&F_3\left(1,-2,1,0,0,0,0,1,0\right),\notag\\
    &F_3\left(1,0,1,1,0,0,0,1,0\right), &&F_3\left(1,1,0,0,1,0,0,1,0\right), &&F_3\left(1,1,1,0,1,0,0,1,0\right),\notag\\
    &F_3\left(1,1,0,0,0,1,0,1,0\right), &&F_3\left(1,0,1,0,0,1,0,1,0\right), &&F_3\left(1,-1,1,0,0,1,0,1,0\right),\notag\\
    &F_3\left(1,-2,1,0,0,1,0,1,0\right), &&F_3\left(0,1,1,0,0,1,0,1,0\right), &&F_3\left(-1,1,1,0,0,1,0,1,0\right),\notag\\
    &F_3\left(-2,1,1,0,0,1,0,1,0\right), &&F_3\left(1,1,1,0,0,1,0,1,0\right), &&F_3\left(1,1,1,-1,0,1,0,1,0\right),\notag\\
    &F_3\left(1,1,1,0,0,1,-1,1,0\right), &&F_3\left(0,0,1,1,0,1,0,1,0\right), &&F_3\left(-1,0,1,1,0,1,0,1,0\right),\notag\\
    &F_3\left(1,0,1,1,0,1,0,1,0\right), &&F_3\left(1,-1,1,1,0,1,0,1,0\right), &&F_3\left(1,-2,1,1,0,1,0,1,0\right),\notag\\
    &F_3\left(1,1,0,0,1,1,0,1,0\right), &&F_3\left(1,1,1,0,1,1,0,1,0\right), &&F_3\left(1,1,0,0,0,0,1,1,0\right),\notag\\
    &F_3\left(1,0,1,0,0,0,1,1,0\right), &&F_3\left(1,1,0,0,1,0,1,1,0\right), &&F_3\left(1,0,1,0,1,0,1,1,0\right),\notag\\
    &F_3\left(1,1,1,0,1,0,1,1,0\right), &&F_3\left(1,1,0,0,0,1,1,1,0\right), &&F_3\left(0,0,1,0,0,1,1,1,0\right),\notag\\
    &F_3\left(1,0,1,0,0,1,1,1,0\right), &&F_3\left(1,-1,1,0,0,1,1,1,0\right), &&F_3\left(1,0,1,-1,0,1,1,1,0\right),\notag\\
    &F_3\left(1,0,1,0,-1,1,1,1,0\right), &&F_3\left(1,0,1,0,0,1,1,1,-1\right), &&F_3\left(0,1,1,0,0,1,1,1,0\right),\notag\\
    &F_3\left(-1,1,1,0,0,1,1,1,0\right), &&F_3\left(-2,1,1,0,0,1,1,1,0\right), &&F_3\left(1,1,1,0,0,1,1,1,0\right),\notag\\
    &F_3\left(1,1,1,-1,0,1,1,1,0\right), &&F_3\left(1,1,1,0,-1,1,1,1,0\right), &&F_3\left(0,0,1,1,0,1,1,1,0\right),\notag\\
    &F_3\left(1,0,1,1,0,1,1,1,0\right), &&F_3\left(1,0,1,1,-1,1,1,1,0\right), &&F_3\left(1,1,0,0,1,1,1,1,0\right),\notag\\
    &F_3\left(1,0,1,0,1,1,1,1,0\right), &&F_3\left(1,0,1,-1,1,1,1,1,0\right), &&F_3\left(1,1,1,0,1,1,1,1,0\right),\notag\\
    &F_3\left(1,1,1,-1,1,1,1,1,0\right), &&F_3\left(1,1,1,0,1,1,1,1,-1\right), &&F_3\left(1,1,1,-2,1,1,1,1,0\right),\notag\\
    &F_3\left(1,1,1,-1,1,1,1,1,-1\right), &&F_3\left(0,0,1,1,0,0,0,0,1\right), &&F_3\left(-1,0,1,1,0,0,0,0,1\right),\notag\\
    &F_3\left(1,0,1,1,0,0,0,0,1\right), &&F_3\left(1,-1,1,1,0,0,0,0,1\right), &&F_3\left(1,0,1,0,0,0,1,0,1\right),\notag\\
    &F_3\left(1,-1,1,0,0,0,1,0,1\right), &&F_3\left(1,-2,1,0,0,0,1,0,1\right), &&F_3\left(0,0,1,0,1,0,1,0,1\right),\notag\\
    &F_3\left(-1,0,1,0,1,0,1,0,1\right), &&F_3\left(0,-1,1,0,1,0,1,0,1\right), &&F_3\left(1,0,0,0,0,1,1,0,1\right),\notag\\
    &F_3\left(1,0,1,0,0,1,1,0,1\right), &&F_3\left(1,-1,1,0,0,1,1,0,1\right), &&F_3\left(1,0,1,-1,0,1,1,0,1\right),\notag\\
    &F_3\left(0,0,0,1,0,1,1,0,1\right), &&F_3\left(1,0,0,1,0,1,1,0,1\right), &&F_3\left(0,0,1,1,0,1,1,0,1\right),\notag\\
    &F_3\left(1,0,1,1,0,1,1,0,1\right), &&F_3\left(0,0,1,0,1,1,1,0,1\right), &&F_3\left(-1,0,1,0,1,1,1,0,1\right),\notag\\
    &F_3\left(1,0,1,0,0,0,0,1,1\right), &&F_3\left(1,-1,1,0,0,0,0,1,1\right), &&F_3\left(0,0,1,1,0,0,0,1,1\right),\notag\\
    &F_3\left(-1,0,1,1,0,0,0,1,1\right), &&F_3\left(1,0,1,1,0,0,0,1,1\right), &&F_3\left(1,-1,1,1,0,0,0,1,1\right),\notag\\
    &F_3\left(1,0,1,1,-1,0,0,1,1\right), &&F_3\left(1,0,1,1,0,-1,0,1,1\right), &&F_3\left(1,-2,1,1,0,0,0,1,1\right),\notag\\
    &F_3\left(1,-1,1,1,-1,0,0,1,1\right), &&F_3\left(1,0,0,0,0,0,1,1,1\right), &&F_3\left(1,0,1,0,0,0,1,1,1\right),\notag\\
    &F_3\left(1,-1,1,0,0,0,1,1,1\right), &&F_3\left(1,-2,1,0,0,0,1,1,1\right), &&F_3\left(1,0,0,0,0,1,1,1,1\right),\notag\\
    &F_3\left(1,0,1,0,0,1,1,1,1\right), &&F_3\left(1,-1,1,0,0,1,1,1,1\right), &&F_3\left(1,0,1,-1,0,1,1,1,1\right),\notag\\
    &F_3\left(0,0,0,1,0,1,1,1,1\right), &&F_3\left(1,0,0,1,0,1,1,1,1\right), &&F_3\left(0,0,1,1,0,1,1,1,1\right),\notag\\
    &F_3\left(1,0,1,1,0,1,1,1,1\right), &&F_3\left(1,-1,1,1,0,1,1,1,1\right), &&F_3\left(1,0,1,1,-1,1,1,1,1\right),\notag\\
    &F_3\left(1,-2,1,1,0,1,1,1,1\right), &&F_3\left(1,-1,1,1,-1,1,1,1,1\right)
\end{align}

\begin{align}
    &F_4\left(1,0,1,0,0,0,1,0,0\right), &&F_4\left(0,0,1,0,0,1,1,0,0\right), &&F_4\left(-1,0,1,0,0,1,1,0,0\right),\notag\\
    &F_4\left(1,0,0,0,0,0,0,1,0\right), &&F_4\left(1,0,1,0,0,0,0,1,0\right), &&F_4\left(1,-1,1,0,0,0,0,1,0\right),\notag\\
    &F_4\left(1,0,1,1,0,0,0,1,0\right), &&F_4\left(1,0,0,0,1,0,0,1,0\right), &&F_4\left(1,0,0,0,0,1,0,1,0\right),\notag\\
    &F_4\left(1,0,1,0,0,1,0,1,0\right), &&F_4\left(1,-1,1,0,0,1,0,1,0\right), &&F_4\left(1,0,1,-1,0,1,0,1,0\right),\notag\\
    &F_4\left(0,0,0,1,0,1,0,1,0\right), &&F_4\left(1,0,0,1,0,1,0,1,0\right), &&F_4\left(0,0,1,1,0,1,0,1,0\right),\notag\\
    &F_4\left(-1,0,1,1,0,1,0,1,0\right), &&F_4\left(0,-1,1,1,0,1,0,1,0\right), &&F_4\left(1,0,1,1,0,1,0,1,0\right),\notag\\
    &F_4\left(1,-1,1,1,0,1,0,1,0\right), &&F_4\left(1,0,1,1,-1,1,0,1,0\right), &&F_4\left(1,0,0,0,1,1,0,1,0\right),\notag\\
    &F_4\left(0,0,0,1,1,1,0,1,0\right), &&F_4\left(1,0,0,1,1,1,0,1,0\right), &&F_4\left(0,0,1,0,0,1,1,1,0\right),\notag\\
    &F_4\left(-1,0,1,0,0,1,1,1,0\right), &&F_4\left(0,0,1,0,1,1,1,1,0\right), &&F_4\left(1,0,1,0,0,0,0,0,1\right),\notag\\
    &F_4\left(1,-1,1,0,0,0,0,0,1\right), &&F_4\left(1,0,1,1,0,0,0,0,1\right), &&F_4\left(1,-1,1,1,0,0,0,0,1\right),\notag\\
    &F_4\left(1,0,1,0,1,0,0,0,1\right), &&F_4\left(1,-1,1,0,1,0,0,0,1\right), &&F_4\left(1,0,1,-1,1,0,0,0,1\right),\notag\\
    &F_4\left(1,0,1,1,1,0,0,0,1\right), &&F_4\left(1,-1,1,1,1,0,0,0,1\right), &&F_4\left(0,0,1,0,1,1,1,0,1\right),\notag\\
    &F_4\left(-1,0,1,0,1,1,1,0,1\right), &&F_4\left(1,0,1,0,0,0,0,1,1\right), &&F_4\left(1,0,1,1,0,0,0,1,1\right),\notag\\
    &F_4\left(1,-1,1,1,0,0,0,1,1\right), &&F_4\left(1,0,1,1,-1,0,0,1,1\right), &&F_4\left(1,0,1,1,0,0,-1,1,1\right),\notag\\
    &F_4\left(1,0,1,0,1,0,0,1,1\right), &&F_4\left(1,-1,1,0,1,0,0,1,1\right), &&F_4\left(1,0,1,1,1,0,0,1,1\right),\notag\\
    &F_4\left(1,0,1,0,0,1,0,1,1\right), &&F_4\left(1,-1,1,0,0,1,0,1,1\right), &&F_4\left(1,0,1,1,0,1,0,1,1\right),\notag\\
    &F_4\left(1,-1,1,1,0,1,0,1,1\right)
\end{align}

\begin{align}
    &F_5\left(1,1,0,0,0,0,0,0,0\right), &&F_5\left(0,1,1,1,0,0,0,0,0\right), &&F_5\left(1,1,0,0,1,0,0,0,0\right),\notag\\
    &F_5\left(0,1,1,0,1,0,0,0,0\right), &&F_5\left(-1,1,1,0,1,0,0,0,0\right), &&F_5\left(0,1,1,1,1,0,0,0,0\right),\notag\\
    &F_5\left(1,1,0,0,0,1,0,0,0\right), &&F_5\left(0,1,1,0,0,1,0,0,0\right), &&F_5\left(-1,1,1,0,0,1,0,0,0\right),\notag\\
    &F_5\left(0,1,0,1,0,1,0,0,0\right), &&F_5\left(1,1,0,1,0,1,0,0,0\right), &&F_5\left(0,1,1,1,0,1,0,0,0\right),\notag\\
    &F_5\left(-1,1,1,1,0,1,0,0,0\right), &&F_5\left(0,1,1,1,0,1,-1,0,0\right), &&F_5\left(1,1,0,0,1,1,0,0,0\right),\notag\\
    &F_5\left(0,1,1,0,1,1,0,0,0\right), &&F_5\left(0,1,0,1,1,1,0,0,0\right), &&F_5\left(1,1,0,1,1,1,0,0,0\right),\notag\\
    &F_5\left(0,1,1,1,1,1,0,0,0\right), &&F_5\left(-1,1,1,1,1,1,0,0,0\right), &&F_5\left(0,1,1,1,1,1,-1,0,0\right),\notag\\
    &F_5\left(0,0,1,0,0,1,1,0,0\right), &&F_5\left(-1,0,1,0,0,1,1,0,0\right), &&F_5\left(1,0,1,0,0,1,1,0,0\right),\notag\\
    &F_5\left(1,-1,1,0,0,1,1,0,0\right), &&F_5\left(0,1,1,0,0,1,1,0,0\right), &&F_5\left(-1,1,1,0,0,1,1,0,0\right),\notag\\
    &F_5\left(1,1,1,0,0,1,1,0,0\right), &&F_5\left(1,1,1,-1,0,1,1,0,0\right), &&F_5\left(1,1,1,0,0,1,1,-1,0\right),\notag\\
    &F_5\left(1,1,0,0,1,0,0,0,1\right), &&F_5\left(1,0,1,0,1,0,0,0,1\right), &&F_5\left(1,-1,1,0,1,0,0,0,1\right),\notag\\
    &F_5\left(1,0,1,-1,1,0,0,0,1\right), &&F_5\left(0,1,1,0,1,0,0,0,1\right), &&F_5\left(-1,1,1,0,1,0,0,0,1\right),\notag\\
    &F_5\left(0,1,1,-1,1,0,0,0,1\right), &&F_5\left(1,1,1,0,1,0,0,0,1\right), &&F_5\left(1,1,1,-1,1,0,0,0,1\right),\notag\\
    &F_5\left(1,1,1,0,1,0,-1,0,1\right), &&F_5\left(1,1,1,-2,1,0,0,0,1\right), &&F_5\left(0,0,1,1,1,0,0,0,1\right),\notag\\
    &F_5\left(-1,0,1,1,1,0,0,0,1\right), &&F_5\left(1,0,1,1,1,0,0,0,1\right), &&F_5\left(1,-1,1,1,1,0,0,0,1\right),\notag\\
    &F_5\left(0,1,1,1,1,0,0,0,1\right), &&F_5\left(-1,1,1,1,1,0,0,0,1\right), &&F_5\left(0,1,1,1,1,-1,0,0,1\right),\notag\\
    &F_5\left(0,1,1,1,1,0,-1,0,1\right), &&F_5\left(0,1,1,1,1,0,0,-1,1\right), &&F_5\left(-2,1,1,1,1,0,0,0,1\right),\notag\\
    &F_5\left(1,1,1,1,1,0,0,0,1\right), &&F_5\left(1,1,1,1,1,-1,0,0,1\right), &&F_5\left(1,1,1,1,1,0,-1,0,1\right),\notag\\
    &F_5\left(1,1,0,0,0,1,0,0,1\right), &&F_5\left(0,1,0,1,0,1,0,0,1\right), &&F_5\left(1,1,0,1,0,1,0,0,1\right),\notag\\
    &F_5\left(0,1,1,1,0,1,0,0,1\right), &&F_5\left(-1,1,1,1,0,1,0,0,1\right), &&F_5\left(0,1,1,1,0,1,-1,0,1\right),\notag\\
    &F_5\left(-2,1,1,1,0,1,0,0,1\right), &&F_5\left(1,1,0,0,1,1,0,0,1\right), &&F_5\left(0,1,1,0,1,1,0,0,1\right),\notag\\
    &F_5\left(-1,1,1,0,1,1,0,0,1\right), &&F_5\left(0,1,0,1,1,1,0,0,1\right), &&F_5\left(1,1,0,1,1,1,0,0,1\right),\notag\\
    &F_5\left(0,1,1,1,1,1,0,0,1\right), &&F_5\left(-1,1,1,1,1,1,0,0,1\right), &&F_5\left(0,1,1,1,1,1,-1,0,1\right),\notag\\
    &F_5\left(0,1,1,1,1,1,0,-1,1\right)
\end{align}

\bibliographystyle{JHEP}
\bibliography{references_inspire.bib,references_local.bib}

\end{document}